\RequirePackage{lineno} 
\documentclass[reprint,twocolumn,showpacs,preprintnumbers,amsmath,amssymb,floatfix, aps,]{revtex4-1}
\usepackage[pdftex]{graphicx}
\usepackage{dcolumn}
\usepackage{bm}
\usepackage{afterpage}
\usepackage{amsmath}
\usepackage{color}
\usepackage{natbib}

\begin{document}

\preprint{APS/123-QED}

\title{Nonlinear magneto-optical resonances for systems with $J\sim$100\newline
observed in K$_{2}$ molecules}

\author{M.~Auzinsh$^1$}
\author{R.~Ferber$^1$}%
\author{I.~Fescenko$^1$}
\email{iliafes@gmail.com}
\author{L.~Kalvans$^1$}%
\author{M.~Tamanis$^1$}%

\affiliation{ $^1$Laser Centre, The University of Latvia, 19 Rainis
Boulevard, LV-1586 Riga, Latvia}%

\date{\today}

\begin{abstract}
We present the results of an experimental as well as theoretical study of nonlinear magneto-optical resonances in diatomic potassium molecules in the electronic ground state with large values of the angular momentum quantum number $J\sim$100. At zero magnetic field, the absorption transitions are suppressed because of population trapping in the ground state due to Zeeman coherences between magnetic sublevels of this state along with depopulation pumping. The destruction of such coherences in an external magnetic field was used to study the resonances in this work.
K$_{2}$ molecules were formed in a glass cell filled with potassium metal at a temperature above 150$^o$C. The cell was placed in an oven and was located in a homogeneous magnetic field \textbf{B}, which was scanned from zero to 0.7 T. \textit{Q}-type and \textit{R}-type transitions were excited with a tunable, single-mode diode laser with central wavelength of 660 nm. Well pronounced nonlinear Hanle effect signals were observed in the intensities of the linearly polarized components of the laser-induced fluorescence (LIF) detected in the direction parallel to the \textbf{B}-field with polarization vectors parallel ($I_{par}$) and perpendicular ($I_{per}$) to the polarization vector of the exciting laser radiation, which was orthogonal to \textbf{B}. 
The intensities of the LIF components were detected for different experimental parameters, such as laser power density and vapor temperature, in order to compare them with numerical simulations that were based on the optical Bloch equations for the density matrix. We report good agreement of our measurements with numerical simulations. Narrow, subnatural line width dark resonances in $I_{per}(B)$ were detected and explained.

\end{abstract}

\pacs{33.57.+c,33.80.Be}

\maketitle

\section{\label{Intro:level1}Introduction}
Magneto-optical effects in atom-laser interactions are due to the excitation of a coherent superposition of the atomic or molecular energy levels with the laser radiation. Light synchronizes the phases of the wave functions of these coherently excited atomic levels. If these states happen to be degenerate Zeeman magnetic sublevels $M_{J}$ of the angular momentum state with angular momentum quantum number $J$, then the application of an external magnetic field \textit{B} can split the Zeeman $M_{J}$ sublevels, removing the degeneracy, and, as a result, destroying the Zeeman coherences of the magnetic sublevels, or, in other words, destroying the phase synchronization of the respective wave functions. These changes of atomic properties due to the destruction of coherences lead to magneto-optical resonances. The first observation of such a magneto-optical resonance in the excited state of Hg atoms was made in 1924 by Wilhelm Hanle~\cite{Hanle}, and now is known as the Hanle effect. Accordingly, one speaks of observing resonances in the Hanle configuration when the magneto-optical resonances appear in a fluorescence signal detected along the magnetic field direction as a function of the applied magnetic field \textit{B} at linearly polarized excitation with the polarization \textbf{E} vector perpendicular to  \textbf{B}. 

If the intensity of light is strong enough, atomic coherences can be created not only in the excited state of an atom, but also in the ground state through nonlinear absorption by applying the optical pumping technique pioneered by Alfred Kastler. These studies led to the Nobel Prize in physics in 1966 being awarded to Alfred Kastler "for the discovery and development of optical methods for studying Hertzian resonances in atoms"~\cite{Kastler}. If these effects are studied in the Hanle configuration, then often these are classified as ground-state Hanle effect. The ground-state Hanle effect in atoms was first observed in~\cite{Lehmann:1964} (see~\cite{Strumia} for a review).

Magneto-optical effects turned out to be such a rich phenomenon, that nowadays they are still actively studied (see, for example, the review papers~\cite{Budker:2002, Alexandrov:2005}). These studies have generated many interesting applications for opto-electronics~\cite{dawes:2005}, medicine~\cite{Groeger:2006} and precise measurement techniques. For example, one of the most sensitive methods of magnetometry is based on exploiting nonlinear magneto-optical effects in an atomic vapor~\cite{Budker:2007}.

Magneto-optical effects have very many manifestations, such as dark and bright 
resonances~\cite{Lukin:1999, Alnis:2001, Dancheva:2000, Alzetta:1976}. 
These resonances occur when laser radiation at zero magnetic field modifies the absorption coefficient of the gas through which it is propagating. The resonance is called dark if the absorption coefficient 
decreases as a result of the interaction with the laser radiation, and bright if the absorption coefficient increases. The laser-induced modification of the absorption coefficient can be eliminated by applying a magnetic field. 
The other manifestations are coherent population trapping~\cite{Arimondo:1996}, electromagnetically induced transparency~\cite{Fleisch:2005} and electromagnetically induced absorption~\cite{Budker:2004}. The phenomenon of the slow light is closely related to these effects as well~\cite{Hau:1999}.

The majority of studies of magneto-optical effects in gases are carried out in atomic media. Although the underlying reason for the appearance of  magneto-optical resonances in molecules is the same as in atoms---Zeeman coherence of magnetic sublevels of a certain angular momentum state---small molecules, 
such as diatomic molecules, offer a new perspective on these effects. 
Molecular states in diatomic molecules typically have properties that differ significantly from the properties atomic states.  

Angular momentum states with $J\sim$100 behave almost classically, which means that angular momentum projection on the quantization axis changes almost continuously. For example, for $J=100$ there are $2J+1=201$ allowed projection values. In contrast to the atomic states with small angular momentum quantum numbers $J$, in molecules angular momentum sometimes can be described even as a classical angular momentum with a well defined orientation in space that changes continuously, see for example~\cite{Molecules}. An exciting perspective can by gained by studying molecular states with small, intermediate and large angular momentum quantum numbers and following how the quantum treatment gradually starts to 
coincide with the classical one.

Another practical reason why the study of molecular states is of particular interest is that they have relatively small magnetic moments, typically $10^{-4}$ to $10^{-5}$ times smaller then in atoms. This allows magnetometry methods based on magneto-optical phenomenon in atoms and applicable to extremely weak magnetic fields, to be used for lager fields. In particular to the magnetic fields of the order of Earth magnetic field.

Another feature that is particular to molecules is that they have negligibly small hyperfine splitting of the ground as well the excited states, unlike alkali atoms which are often used to study magneto-optical effects. This can simplify the analysis of the interaction between laser radiation and the molecules, because we do not need to consider interaction of radiation with several, partially overlapping hyperfine transitions simultaneously.

In order for bright resonances to form, it is necessary that a system -- atom or molecule -- cycles several times between the excited state and the initial state~\cite{Arimondo:1996, Papoyan:2002, auzinsh_book}. 
However, light absorption in diatomic molecules is usually described in terms of absorption in an open system: the molecule rarely returns to its initial rovibronic state after excitation by a photon, because spontaneous transitions from the excited state are allowed to a very large number of rovibronic levels in the ground state. Nevertheless there still is a small probability for the molecule to return to the initial state. Therefore, one of the tasks for this study is to determine, using theoretical and experimental means, if sufficient cycles between excited and initial state can take place in diatomic molecules for bright resonances to be observable.

\begin{figure}[htbp]
    \centering
         \resizebox{\columnwidth}{!}{\includegraphics{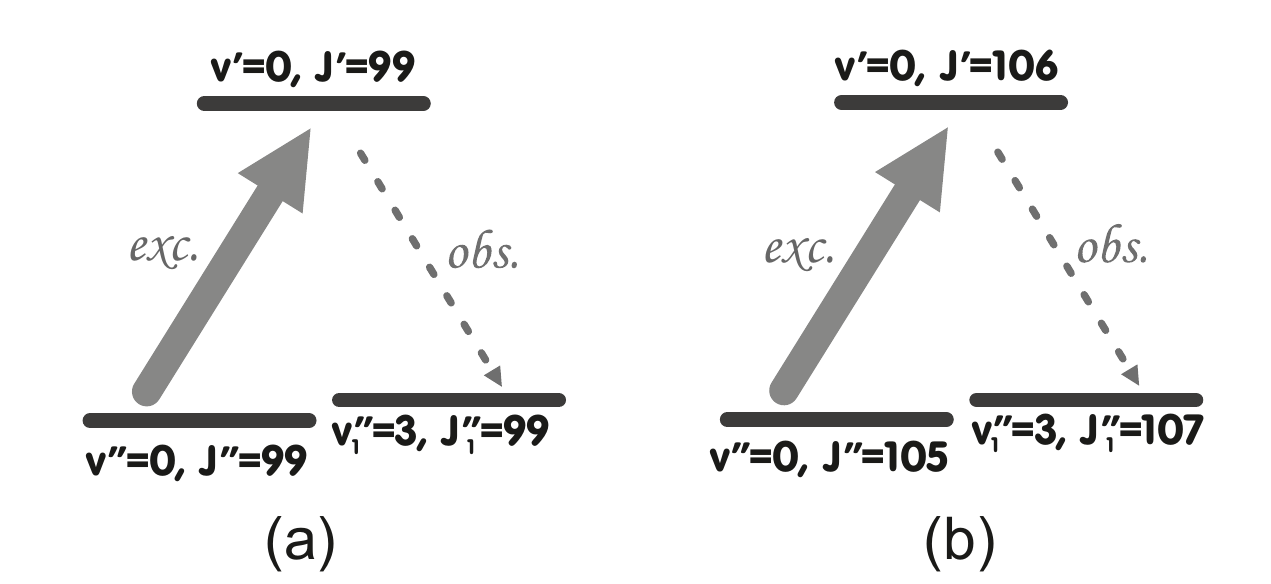}}
    \caption{\label{fig:levels} Excitation and observation scheme for the transitions used in this study: (a) \textit{Q}-type transition with $\Delta J=J'-J''=0$ and excitation frequency 15111.270 cm$^{-1}$; (b) \textit{R}-type transition with $\Delta J=J'-J''=+1$ and excitation frequency 15111.940 cm$^{-1}$. In the latter case, \textit{P}-type transitions were used for observations.}
\end{figure}
To our knowledge, the first mention of an increase in the intensity of the laser-induced fluorescence (LIF) of diatomic molecules when a magnetic field ($B\geq 1$ T) was applied was reported in~\cite{Broyer:1973} and~\cite{Solarz:1972} for the I$_{2}$ molecule. An increase in the LIF intensity of up to $30\%$  was observed under strong Ar$^{+}$ or Kr$^{+}$ laser excitation of about 1-10 W/cm$^{2}$ at different orientations of the \textbf{E}-vector of linearly polarized light. The authors did not attribute the result to the ground-state Hanle effect, because it was observed also at \textbf{E} parallel to \textbf{B}. The first direct experimental studies of the ground-state Hanle signal were reported in~\cite{Ferber:1979} for the diatomic molecules K$_{2}$ and Na$_{2}$ in their X$^{1}\Sigma^{+}_{g}$ electronic ground states with large angular momenta $J''=73$ and 99, respectively. Changes of the degree of LIF polarization with increasing \textbf{B} perpendicular to \textbf{E} have been observed in conditions of nonlinear \textit{Q}-type excitation $(J'-J''=0)$ with a fixed frequency multimode laser that accidentally excited a particular B$^{1}\Pi$ ($v'$, $J'$) rovibronic level in these systems. Under such conditions, the so-called, ground-state "optical pumping", first observed in~\cite{Drullinger:1969} for Na$_{2}$, leads to a diminished degree of LIF polarization at $B=0$. In particular, in~\cite{Ferber:1979} for the K$_{2}$ molecule, LIF was excited by the 632.8 nm line of a multimode He-Ne laser with a power density of about 1 W/cm$^{2}$; the nonlinear ground-state Hanle effect manifested itself as an increase in the degree of LIF polarization of about $5\%$ as the magnetic field \textit{B} was increased up to 0.8 T. At larger \textit{B} values LIF polarization diminished because of the excited B$^{1}\Pi$ state Hanle effect. 

The nonlinear Hanle signals in diatomic molecules observed in~\cite{Ferber:1979} (see also~\cite{Molecules} for a review), were analyzed using the classical treatment developed for the asymptotic limit   $J\rightarrow \infty$, by Ducloy~\cite{Ducloy:1976} in terms of the balance equations for the classical probability density, accounting for the precession of classical angular momentum \textbf{J} in the magnetic field. The connection of such an approach to Zeeman coherences has been discussed in~\cite{Molecules}.  

To explain these results, a model was developed applying an expansion of the density matrix via polarization (multipole) moments (see~\cite{Molecules} for a review), which made it possible to interpret the additional narrow structure observed in the vicinity of \textit{B}=0 of a nonlinear Hanle signal~\cite{Ferber:1979} as a manifestation, via a fourth-rank moment, of the sixth-rank polarization moment of the ground state $J\gg1$~\cite{Molecules}.  A comparison of the quantum mechanical and classical treatments of the nonlinear Hanle signals based on the balance equations is given in~\cite{Auzinsh:1991, Molecules}. 

The motivation of the present study is to revisit the nonlinear ground state Hanle effect of a diatomic molecule with $J\sim100$ and to study in detail the peculiarities observed in earlier works~\cite{Ferber:1979, Molecules}. We intended to achieve a well controlled excitation by using the  narrow line of a single-mode, scanned, diode laser and to examine the nonlinear magneto-optical resonances under various excitation power densities and relaxation conditions, as well as to apply the detailed theoretical treatment proposed in~\cite{Renzoni:2001a, AlnisJPB:2001, Papoyan:2002}. Such a treatment already has been applied successfully to atoms; it included averaging over the Doppler profile and allowed reproducing magneto-optical signals with experimental accuracy at different laser frequency detuning (see, for instance,~\cite{Auzinsh:2009} and references therein).  

\section{\label{Experiment:level1}Experiment}
K$_{2}$ molecules were formed in an 8 cm long custom-made glass cell with a side-arm, which contained natural mixture of potassium isotopes. The cell had a diameter of 16 mm and the windows on both ends were flat and parallel. 
The cell was partly blackened from the outside to reduce reflections.  It was connected by a dry valve to a vacuum system in order to pump away impurities that diffused from the walls of the cell. The cell was placed in an oven that was located between the poles of an electromagnet, which were separated by a  5 cm gap. The oven was made from non-magnetic materials and contained two separately powered heating elements. The temperature of the side-arm varied between 150$^\circ$C and 185$^\circ$C. The temperature in the cell was maintained \textcolor{red}{} 30$^\circ$C higher to avoid metal deposition on the walls of 
the cell. 

\begin{figure}[htbp]
    \centering
         \resizebox{\columnwidth}{!}{\includegraphics{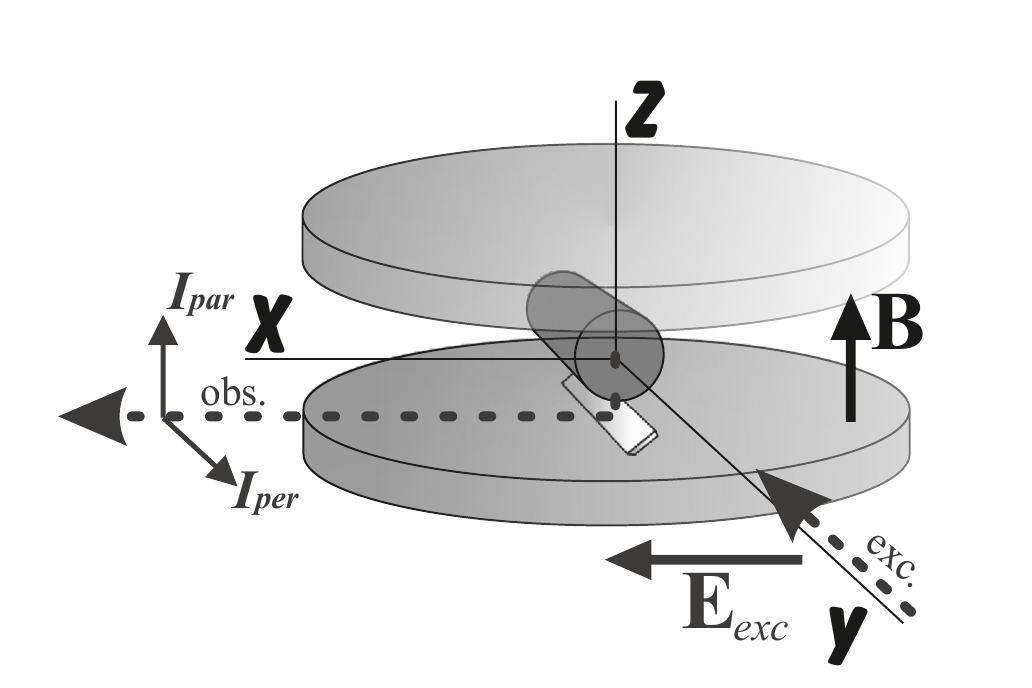}}
    \caption{\label{fig:setup} Experimental geometry. The relative orientation of the laser beam (exc), its polarization vector $\textbf{E}_{exc}$, magnetic field vector \textbf{B}, and observation direction (obs) are shown.}
\end{figure}
The \textit{Q}-type and \textit{R}-type X$^{1}\Sigma_{g}^{+}\rightarrow$ B$^{1}\Pi_{u}$ transitions (see Fig.~\ref{fig:levels}) in the K$_{2}$ molecule were excited at frequencies of 15111.270 cm$^{-1}$ and 15111.940 cm$^{-1}$, respectively, with a tunable, single-mode, external-cavity, cw diode laser with a \textit{Mitsubishi} ML101J27 laser diode whose central wavelength was 660 nm. Its temperature and current were stabilized by \textit{Thorlabs} controllers. The laser beam passed through a Glan-Thompson prism to ensure a high degree of linear polarization. The spectral width of the laser line was assumed to be below 10 MHz. The diameter of the laser beam was defined as the full width at half maximum (FWHM) of the laser beam intensity distribution, and the laser power density was defined as the total laser power divided by the cross-sectional area of the laser beam, as calculated with this beam diameter. The laser beam diameter was 1.1 mm; its cross-sectional area was measured  by a \textit{Thorlabs} BP 104-VIS beam profiler. The maximal laser power at the entrance of the cell was 22 mW. The laser frequency was monitored with a \textit{HighFinesse} WS-7 wavemeter with an relative accuracy of 10 MHz. The B$^{1}\Pi_{u}\rightarrow$X$^{1}\Sigma_{g}^{+}$ LIF spectra were investigated with an ISF-125 high resolution Fourier transform spectrometer \textit{Bruker Optik GmbH} in order to select the proper excitations. 

The LIF from a 1.5 cm long region of the cell was collected by a two-lens system and imaged onto the input slit of a DFS-12 double monochromator with an inverse dispersion of 0.5 nm/mm. The monochromator had selected LIF transitions  15023.476 cm$^{-1}$ in case of the \textit{Q}-excitation and 15001.650 cm$^{-1}$ in case of the \textit{R}-excitation. In order to suppress a nearby line from another level that was simultaneously excited within Doppler width, it was necessary to detune the laser frequency from the resonance by $+200$ MHz for \textit{Q}-excitation and by $+170$ MHz for \textit{R}-excitation. The LIF intensities were detected by a \textit{Hamamatsu} photomultiplier module via an \textit{Electron Tubes} CT2 time counter. The electromagnet, powered by two \textit{Agilent} N5770A DC power supplies connected in parallel, was used to produce a homogeneous magnetic field \textit{B} up to 0.7 T. The magnetic field was calibrated by a digital teslameter with an accuracy of $5\cdot10^{-4}$ T. Two switchable sheet polarizers were inserted in front of the slit to measure the LIF intensity that was polarized either parallel (\textit{I}$_{par}$ ) or orthogonal (\textit{I}$_{per}$ ) to the exciting laser polarization vector \textbf{E} in the Hanle configuration. The experimental geometry is shown in Fig.~\ref{fig:setup}, which depicts the relative orientation of the laser beam, the polarization vector of the laser light $\textbf{E}_{exc}$, the magnetic field \textbf{B}, and the observation direction. To observe the fluorescence in the direction along the magnetic field, a mirror was attached to one of the poles of the electromagnet. 

The fluorescence was recorded while the magnetic field was scanned, in discrete steps, from 0 to 0.7 Tesla in one direction. At each step, the signal was collected during 2 seconds. Thus, a typical scan lasted 40 seconds, including the time necessary for the establishment of the constant magnetic field. Usually, from 40 to 80 scans were performed and averaged.  The dependences of the LIF intensities (\textit{I}$_{par}$ and \textit{I}$_{per}$) on the magnetic field were detected separately and then normalized to the \textit{I}$_{par}$ value at $B=0$.  The degree of polarization was calculated as $P=(I_{par} -I_{per})/(I_{par}+I_{per})$, accounting for different losses in the detection system for each component of the LIF polarization.

\section{\label{Theory:level1}Theoretical model}
A theoretical model had been developed previously in order to describe bright and dark resonances of atomic alkali metals in vapor cells~\cite{Blushs:2004, Auzinsh:2009}, and a detailed description of this theoretical model can be found in these references. We give a brief summary of the model here for convenience and emphasize the changes made in order to treat diatomic molecules. The model describes the internal molecular dynamics by a density matrix $\rho$, which also depends parametrically on the classical coordinates of the molecular center of mass. The time evolution of the density matrix $\rho$ follows the optical Bloch equations (OBEs)~\cite{Stenholm:2005}:
\begin{equation}
        i\hbar \frac{\partial \rho}{\partial t} = \left[\hat{H},\rho \right] + i \hbar\hat{R}\rho.
        \label{obe}
\end{equation}
The Hamiltonian $\hat{H}$ is assumed to be in the form $\hat{H} = \hat{H}_0 + \hat{H}_B + \hat{V}$. Here, $\hat{H}_0$ describes the unperturbed molecular energy structure, $\hat{H}_B$ describes the molecule's interaction with the static magnetic field, and $\hat{V} = -\hat{\textbf{\textit{d}}} \cdot \textbf{\textit{E}}(t)$ is the dipole interaction operator that describes the interaction between the molecule and the exciting radiation. The relaxation operator $\hat{R}$ includes the spontaneous emission rate, which equals the spontaneous decay rate of the excited state $\Gamma$, and the combined transit and collision relaxation rate $\gamma$, which can be estimated from the exciting laser's beam diameter, the temperature and density of the gas in the cell, as well as the effective cross-section of the collisions; more details on this evaluation are given in the next section. 

The exciting radiation is described classically as an oscillating electrical field with a stochastically fluctuating phase, which results in a Lorentzian frequency distribution with full width at half maximum (FWHM) $\Delta\omega$. The central frequency changes as the classically moving molecules experience the Doppler shift. Next, the rotating wave approximation~\cite{Allen:1975} is employed to obtain a system of stochastic differential equations from~\eqref{obe}. Only the equations describing the time evolution of the optical coherences contain the stochastic phase variables. Accounting for the fact that the experimentally observed light intensity is averaged over time intervals that are large compared to the characteristic time of the phase fluctuations, the optical coherences are adiabatically eliminated from the system of differential equations. The equations corresponding to the optical coherences are integrated~\cite{Stenholm:2005}, while the stochastic phase variable is decorrelated and formally averaged ~\cite{Kampen:1976}, and the obtained results are substituted in equations corresponding to the Zeeman coherences~\cite{Auzinsh:2009}. The method yields rate equations for the Zeeman coherences that are valid for either spectrally broad exciting radiation $(\Delta\omega \gg \Gamma)$ or stationary excitation conditions:
\begin{align}
\dfrac{\partial\rho_{g_ig_j}}{\partial t} =&\bigl(\Xi_{g_ie_m} + \Xi_{e_kg_j}^{\ast}\bigr)\underset{e_k, e_m
}{ \sum }\bigl(d_1^{g_ie_k}\bigr)^{\ast}d_1^{e_mg_j}\rho_{e_ke_m} \nonumber \\
&- \underset{e_k,g_m}{\sum }\Bigl[\Xi_{e_kg_j}^{\ast} \bigl(d_1^{g_ie_k}\bigr)^{\ast}d_1^{e_kg_m}\rho_{g_mg_j} \nonumber \\
&+\Xi_{g_ie_k} \bigl(d_1^{g_me_k}\bigr)^{\ast}d_1^{e_kg_j}\rho_{g_ig_m}\Bigr] - i\omega_{g_ig_j}\rho_{g_ig_j} \nonumber \\
&+ \alpha\underset{e_k, e_m}{\sum}\Gamma_{g_ig_j}^{e_ke_m}\rho_{e_ke_m} -\gamma\rho_{g_ig_j} + \lambda\delta\bigl(g_i, g_j\bigr)
\label{rate1}
\end{align}
and
\begin{align}
\dfrac{\partial\rho_{e_ie_j}}{\partial t} =&\bigl(\Xi_{e_ig_m}^{\ast} + \Xi_{g_ke_j}\bigr) \underset{g_k, g_
m}{\sum }d_1^{e_ig_k}\bigl(d_1^{g_me_j}\bigr)^{\ast}\rho_{g_kg_m} \nonumber \\
&- \underset{g_k,e_m}{\sum }\Bigl[\Xi_{g_ke_j} d_1^{e_ig_k}\bigl(d_1^{g_ke_m}\bigr)^{\ast}\rho_{e_me_j} \nonumber \\
&+\Xi_{e_ig_k}^{\ast} d_1^{e_mg_k}\bigl(d_1^{g_ke_j}\bigr)^{\ast}\rho_{e_ie_m}\Bigr]  \nonumber \\
&- i\omega_{e_ie_j}\rho_{e_ie_j} - \Gamma\rho_{e_ie_j}
\label{rate2}.
\end{align}
In these equations $\rho_{g_ig_j}$ and $\rho_{e_ie_j}$ are the density matrix elements for the ground and excited states, respectively. Quantities $\Xi_{g_ie_j}$ and $\Xi_{e_ig_j}^{\ast}$ describe the coupling between the ground and excited states induced by the laser field; they depend on the Rabi frequency ($\Omega_R$), which is discussed in the next section, as well as on the spontaneous decay rate of the excited state ($\Gamma$), the exciting laser's linewidth ($\Delta\omega$) and the actual frequency detuning between particular transition of two magnetic sublevels and the excitation frequency:
\begin{equation}\label{pumping}
\Xi_{ij} = \frac{\Omega_R^2}{\frac{\Gamma+\gamma+\Delta\omega}{2}+\dot\imath\left(\bar\omega-\mathbf{k}_{\bar\omega}\mathbf{v}+\omega_{g_ie_j}\right)}.
\end{equation}
The complex part in the denominator of \eqref{pumping} includes the laser detuning from the exact optical transition frequency ($\bar\omega$), the Doppler shift ($\mathbf{k}_{\bar\omega}\mathbf{v}$), and the Zeeman effect ($\omega_{g_ie_j}$).

The dipole transition matrix elements $d_1^{e_ig_j}$ between the ground state $i$ and the excited state $j$ can be calculated according to angular momentum theory in general and the Wigner-Eckart theorem in particular~\cite{auzinsh_book}. The energy splittings between the magnetic sublevels of either the ground or the excited state are denoted by $\omega_{ij}$; within the scope of the present research the shifts are assumed to be caused by the linear Zeeman effect. We assume our transition to be open (i.e., not all of the molecules from the excited state decay to the initial ground state), and the recurrence is described by the coefficient 
$\alpha$; $\Gamma_{g_ig_j}^{e_ie_j}$ describes the coherence transfer via spontaneous transitions from the excited to the ground state. Re-population of the ground state due to non-optical processes, such as unpolarized molecules in which coherence of magnetic sublevels has not been created flying into the interaction region or molecules loosing their coherence due to collisions, occurs with rate $\lambda$ [$\delta\bigl(g_i, g_j\bigr)$ is a Kronecker delta function]. We assumed that the molecular equilibrium density outside the interaction region is normalized to unity, and so $\lambda = \gamma$.

The experiments are assumed to take place under stationary excitation conditions, so $\partial\rho_{g_ig_j}/\partial t = \partial\rho_{e_ie_j}/\partial t = 0$. Thus, one can reduce the differential equations \eqref{rate1}~+~\eqref{rate2} to a system of linear algebraic equations, which, when solved, yields the density matrices for the molecular ground and excited states. The linear algebraic system can be solved analytically for small rotational angular momentum $J$ values, but the complexity of the problem increases rapidly when the angular momentum is increased: the number of equations in the linear system is proportional to $(2J+1)^2$. Thus, for the $J$ values discussed in the present study, the system \eqref{rate1}~+~\eqref{rate2} can be solved only numerically for particular excitation conditions as the number of equations in the system is on the order of $10^5$. If the linear algebraic system is written in matrix form $A\mathbf{x} = \mathbf{b}$ one needs to diagonalize the matrix of coefficients $A$ in order to solve the linear system. The fact that most of the elements in the matrix of coefficients are equal to zero is employed to optimize the process of diagonalization and thus the use of computational resources. A set of routines (UMFPACK) based on the Unsymmetric MultiFrontal method~\cite{Davis2004_1} is used to perform decomposition of the sparse matrix of the linear system into a product of a lower and an upper triangular matrix (LU factorization). 

\begin{figure}[htbp]
    \centering
       \resizebox{\columnwidth}{!}{\includegraphics{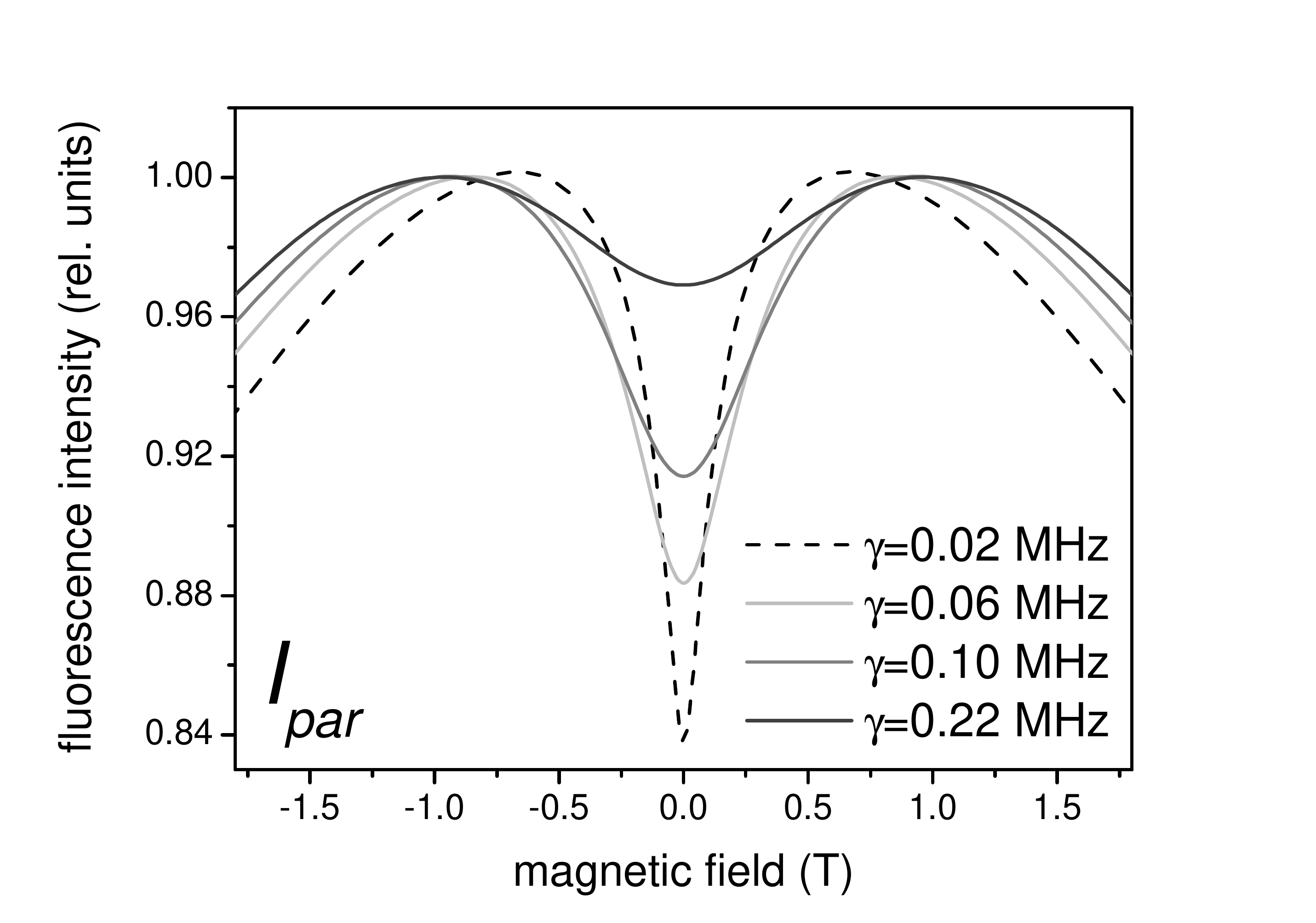}}
       \caption{\label{fig:Theory} Simulation of the parallel polarization component of the LIF intensity versus the magnetic field for K$_{2}$ \textit{Q}-type excitation X$^{1}\Sigma_{g}^{+}\rightarrow$B$^{1}\Pi_{u}$ with $J'=J''=99$, at $\Omega_{R}$=27 MHz for different relaxation rates $\gamma$.}
\end{figure}

Once the density matrices are known, one obtains the observed fluorescence intensity as follows:
\begin{equation}
        I_f(\tilde{\textbf{e}})=\tilde{I}_0\underset{g_i, e_i, e_j}{\sum}d^{(ob)*}_{g_ie_j}d^{(ob)}_{e_ig_i}\rho_{e_ie_j},
        \label{fluorescence}
\end{equation}
where $\tilde{I}_0$ is a constant of proportionality. Two orthogonal polarization components (see Fig. \ref{fig:setup}) are calculated. To describe the influence of the Doppler effect, the fluorescence is summed over the different atomic velocity groups. For the averaging over the frequency, the step size is chosen to be smaller than the natural linewidth of the transition. Figure~\ref{fig:Theory} illustrates the calculated LIF signals of one of polarization components obtained in this way using parameters which were typical for the experiment performed. The following parameters were used for the K$_{2}$ molecule: the spontaneous decay time $\Gamma^{-1}=12\cdot10^{-9}$ s~\cite{Ferber:1979} for the B$^{1}\Pi_{u}$ state, the Land\'e factor of the excited state g$_{J'}=1/J'(J'+1)$~\cite{Herzberg}; the Land\'e factor of the ground state g$_{J''}=(1.01\pm 0.04)\cdot10^{-5}$~\cite{Molecules}. It can be seen that the present theoretical model predicts well pronounced dark resonances in the K$_{2}$ molecule. The analysis demonstrated that the bright resonances should not been observed for such an open system. 

\section{\label{Results:level1}Results and Discussion}
\begin{figure}[htbp]
    \centering
       \resizebox{\columnwidth}{!}{\includegraphics{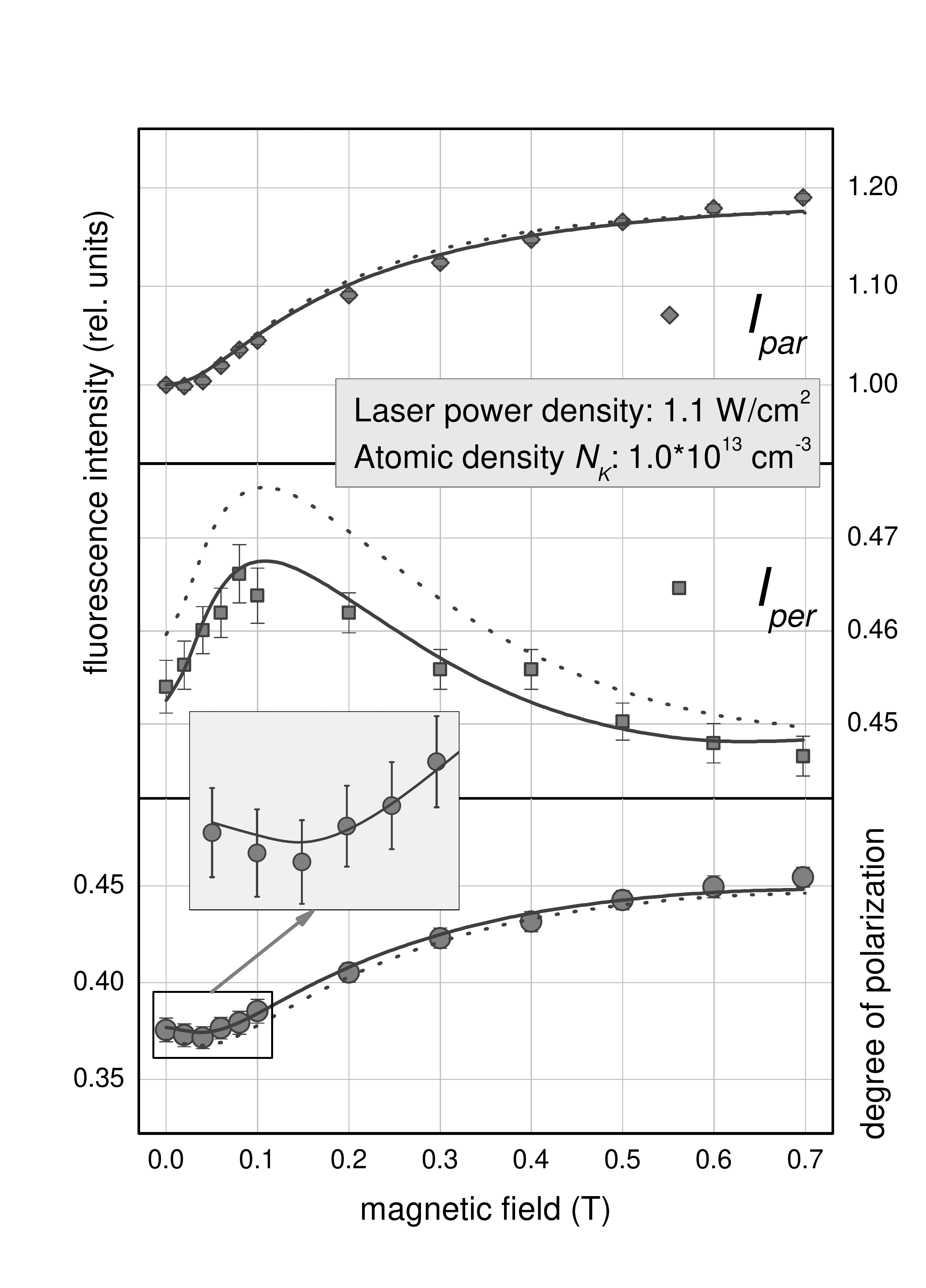}}
       \caption{\label{fig:Q5} Experimental (scattered dots) LIF intensities $I_{par}$, $I_{per}$ and degree of polarization \textit{P} versus the magnetic field  observed in K$_{2}$ for \textit{Q}-type excitation with a side-arm temperature of 150$^\circ$C.  The solid lines are the result of simulations at $\Omega_{R}$=27 MHz and $\gamma$=0.03 MHz, and with a 200 MHz detuning of the excitation frequency from the resonance. 
The dotted lines are calculated for the same parameters, but without the laser frequency detuning.}
\end{figure}

\begin{figure}[htbp]
    \centering
       \resizebox{\columnwidth}{!}{\includegraphics{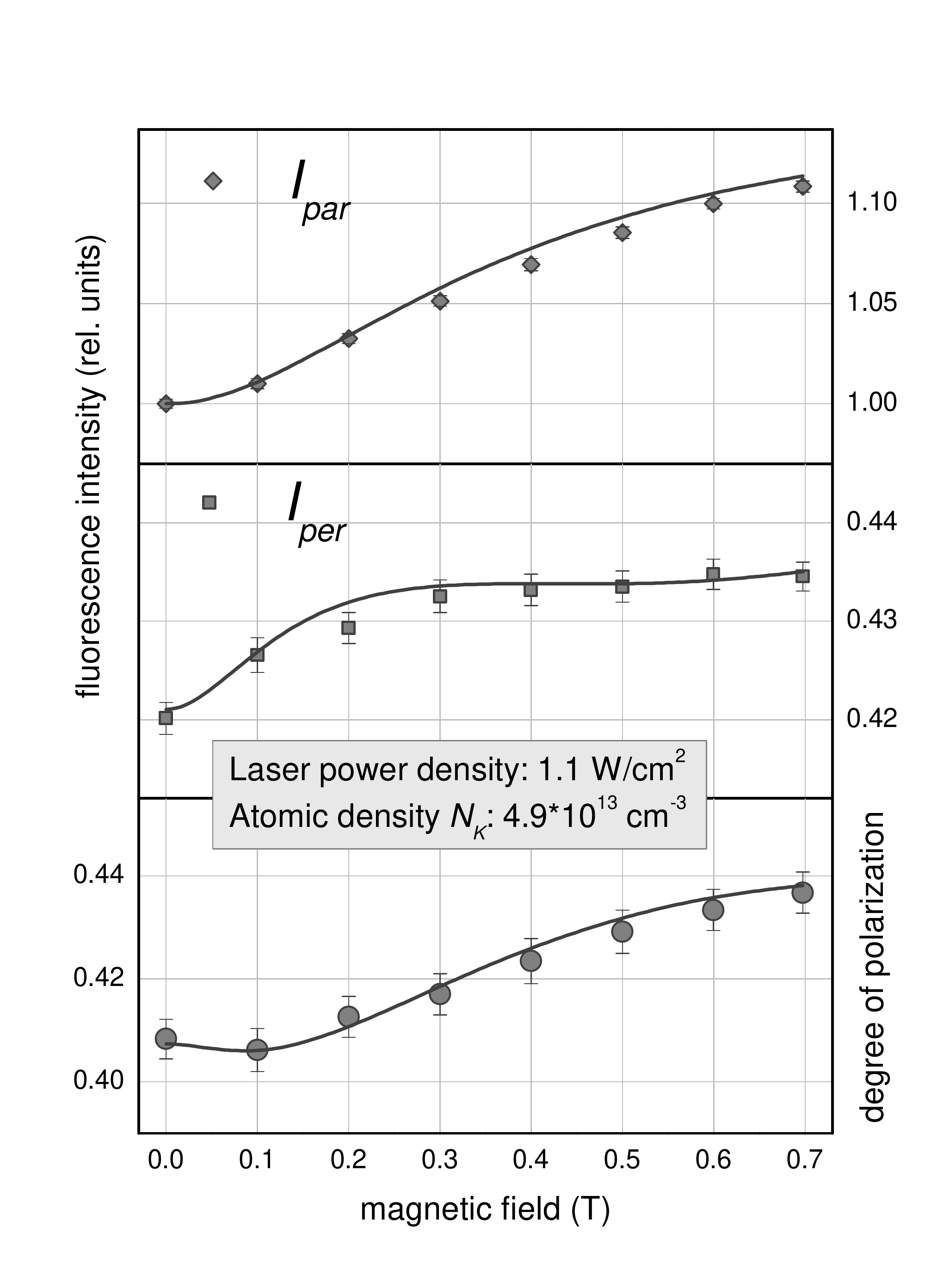}}
       \caption{\label{fig:Q3} Experimental (scattered dots) LIF intensities $I_{par}$, $I_{per}$ and degree of polarization \textit{P} versus magnetic field  observed in K$_{2}$ for \textit{Q}-type excitation with a side-arm temperature of 180$^\circ$C.  The solid lines are the results of a simulation at $\Omega_{R}$=27 MHz and $\gamma$=0.08 MHz, with a 200 MHz detuning of the excitation frequency from the resonance.}
\end{figure}

\begin{figure}[htbp]
    \centering
       \resizebox{\columnwidth}{!}{\includegraphics{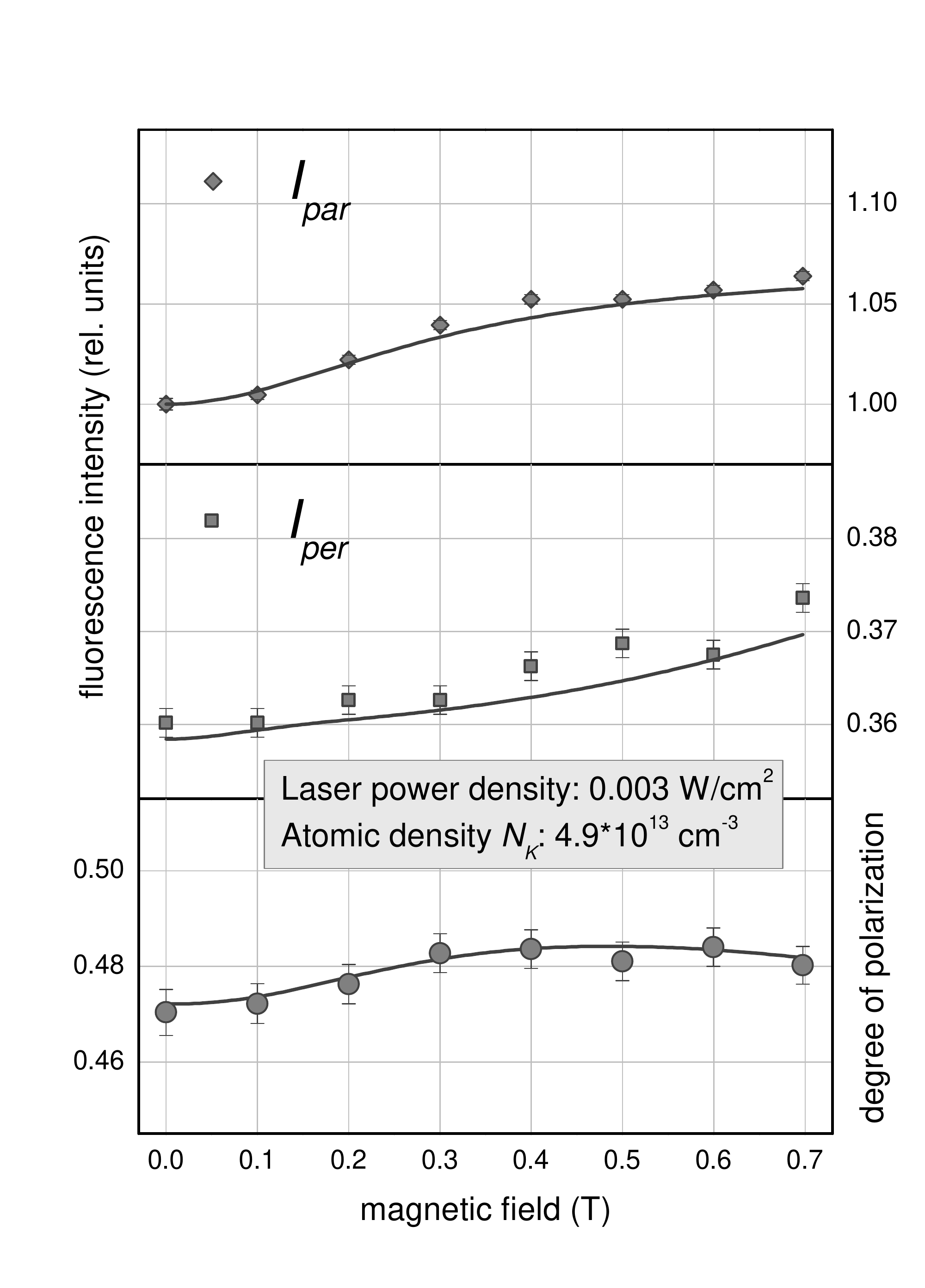}}
       \caption{\label{fig:Q4} Experimental (scattered dots) LIF intensities $I_{par}$, $I_{per}$ and degree of polarization \textit{P} versus magnetic field observed in K$_{2}$ for \textit{Q}-type excitation at reduced laser power density (the temperature in the side-arm was 180$^\circ$C).  The solid lines are the result of a simulation at $\Omega_{R}$=10 MHz and $\gamma$=0.09 MHz, with a 200 MHz detuning of the excitation frequency from the resonance.}
\end{figure}

\begin{figure}[htbp]
    \centering
       \resizebox{\columnwidth}{!}{\includegraphics{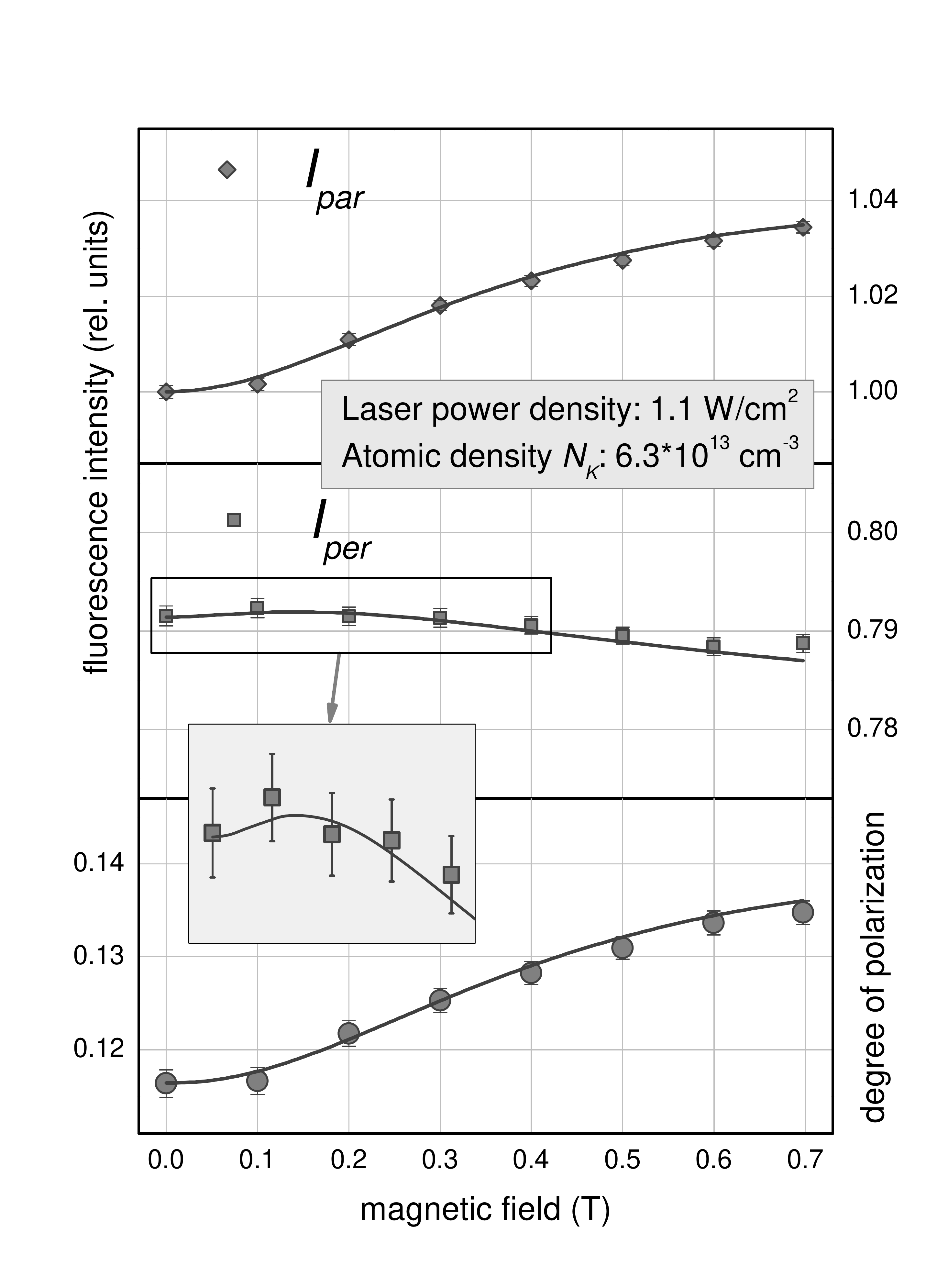}}
       \caption{\label{fig:R2} Experimental (scattered dots) LIF intensities $I_{par}$, $I_{per}$ and degree of polarization \textit{P} versus magnetic field  observed in K$_{2}$ for \textit{R}-type excitation with a side-arm temperature of 185$^\circ$C. The solid lines are the result of a simulation at $\Omega_{R}$=27 MHz and $\gamma$=0.09 MHz, with a 170 MHz detuning of the excitation frequency  from the resonance.}
\end{figure}

The measured dependences on the magnetic field of the LIF intensities and the degree of polarization for the \textit{Q}-type excitation are presented in Figs.~\ref{fig:Q5}--\ref{fig:Q4} for different combinations of potassium vapor densities and laser power densities. This set of data is complemented by an example of the \textit{R}-type excitation presented in Fig.~\ref{fig:R2}. The figures show stacked plots of the intensities \textit{I}$_{par}(B)$ and \textit{I}$_{per}(B)$ and of the degree of polarization $P(B)$. The intensities are given in relative units, and the value of \textit{I}$_{par}$ at \textit{B} = 0 is normalized to unity. The scattered dots represent the experimental data, while the solid lines represent calculations based on Eqs.\eqref{rate1}~--~\eqref{fluorescence}.  

The laser power density and the number density of the potassium atoms $N_{K}$ are given in each figure. The number density $N_{K_{2}}$ of the K$_{2}$ molecules is by about two orders of magnitudes less than $N_{K}$. The density of the potassium atoms and molecules is calculated from the temperature of the metal vapor according to~\cite{Nesmeyanov}. The values of the Rabi frequency $\Omega_{R}$ and of the ground-state relaxation rate $\gamma$ can be roughly estimated from experimental parameters. The Rabi frequency is proportional to the excitation probability of the transition and the laser field amplitude \textbf{E}~\cite{auzinsh_book}:
\begin{equation}
	\Omega_R \propto \vert \textbf{E} \vert \propto \sqrt{I}.
\label{Rabi}
\end{equation}
A simple way to estimate the coefficient of proportionality in \eqref{Rabi} is by calculating the saturation laser power density $I_{sat}$ at which the Rabi frequency becomes equal to the rate of spontaneous emission (assuming that $\gamma \ll \Gamma$) as is shown in~\cite{Alnis:2003}.  

The ground-state relaxation rate $\gamma$ can be described by a sum of the collision-induced relaxation rate $\gamma_{col}$ and the transit relaxation rate $\gamma_{tran}$, which is caused by molecules escaping laser beam region and ``fresh" molecules flying into it:
\begin{equation}
\gamma=\gamma_{col}+\gamma_{tran}=\sigma_{col} N_{K}\langle\textit{\textbf{v}}\rangle+\dfrac{\langle\textit{\textbf{v}}\rangle}{d_{L}},
\label{gamma}
\end{equation}
where $d_{L}$ is the laser beam diameter, $\langle\textit{\textbf{v}}\rangle$ is the mean thermal velocity and $\sigma_{col}$ is the K$_{2}($X$^{1}\Sigma_{g}^{+})+$K inelastic collision cross section. 

Figure~\ref{fig:Q5} shows the LIF signal for \textit{Q}-type excitation with the temperature in the side-arm of the cell held as low as reasonably possible in order to reduce the collision-induced relaxation. The manifestation of the nonlinear ground-state Hanle effect, or dark resonance, is readily visible with a contrast of about 0.18 in \textit{I}$_{par}(B)$ as predicted by calculations (solid line) at $\gamma=0.03$ MHz which agrees well with its estimate by \eqref{gamma} (see also Figure~\ref{fig:Theory}). 
The \textit{I}$_{per}$ dependence on \textit{B} is more complicated. It shows a dark resonance that is noticeably narrower than in \textit{I}$_{par}(B)$ with a smooth decline at \textit{B} $>$ 0.1 T. Such \textit{I}$_{par}(B)$ and \textit{I}$_{per}(B)$ dependences yield the polarization degree $P(B)$ as shown in the lowest graph in Fig.~\ref{fig:Q5}. Note that the degree of polarization $P$ at $B=0$ is about 0.37, which is markedly smaller than the limit at weak excitation, close to 1/2~\cite{Molecules}. A peculiarity in the form of a peak at $P(0)$ can be distinguished (see inset). Such additional structure in $P(B)$ was discovered in~\cite{Ferber:1979} (see~\cite{Molecules} for a review). The narrow dark resonance in \textit{I}$_{per}(B)$ was not revealed in~\cite{Ferber:1979} since only the resonances in $P(B)$ were studied.  

In order to test how the detuning could affect the signal, we calculated numerically the LIF dependence on magnetic field for an excitation without any detuning from the resonance frequency. The results are shown in Fig.~\ref{fig:Q5} by dotted lines. As can be seen, such detuning does not significantly affect the shapes of the magneto-optical resonances.

Figure~\ref{fig:Q3} shows the resonance signals at \textit{Q}-type excitation at the same laser power density as in Fig.~\ref{fig:Q5} but at higher temperature in the side-arm. Compared to the situation in Fig.~\ref{fig:Q5}, the contrast of the resonance in \textit{I}$_{par}(B)$ is smaller and the degree of polarization at $B=0$ increases up to $P(0)=0.42$. It can be seen that the resonance becomes broader, which is caused by the increase of $\gamma$ due to the growth of $\gamma_{col}$ with $N_{K}$ (see expression~\eqref{gamma}). When increasing the temperature from 150$^\circ$C to 180$^\circ$C, the value of $N_{K}$ increases by about 5 times (see Figures~\ref{fig:Q5},~\ref{fig:Q3}), and the collisional process becomes dominant since $\gamma$ increases from 0.03 to 0.08 MHz, which significantly exceeds the transit relaxation rate. The change in the shape of the dark resonance is most pronounced in the \textit{I}$_{per}(B)$ signal. 

We also studied how the resonance shape varies as a function of laser power density. Figure~\ref{fig:Q4} shows the LIF intensities and degree of polarization for \textit{Q}-type excitation at much lower laser power density that was about 360 times weaker than in Figs.~\ref{fig:Q5}--\ref{fig:Q3}. Under these conditions the contrast of the resonance in \textit{I}$_{par}(B)$, \textit{I}$_{per}(B)$ and $P(B)$ was markedly diminished, and the degree of polarization at $B=0$ approached its maximum value at weak excitation close to 1/2.

Figure~\ref{fig:R2} shows the LIF intensities and degree of polarization for \textit{R}-type excitation. The structure in the curves was similar to the structure observed for \textit{Q}-type excitation, which indicates that the same mechanism was mainly responsible for the structure in both types of excitation.

\begin{figure}[htbp]
    \centering
       \resizebox{\columnwidth}{!}{\includegraphics{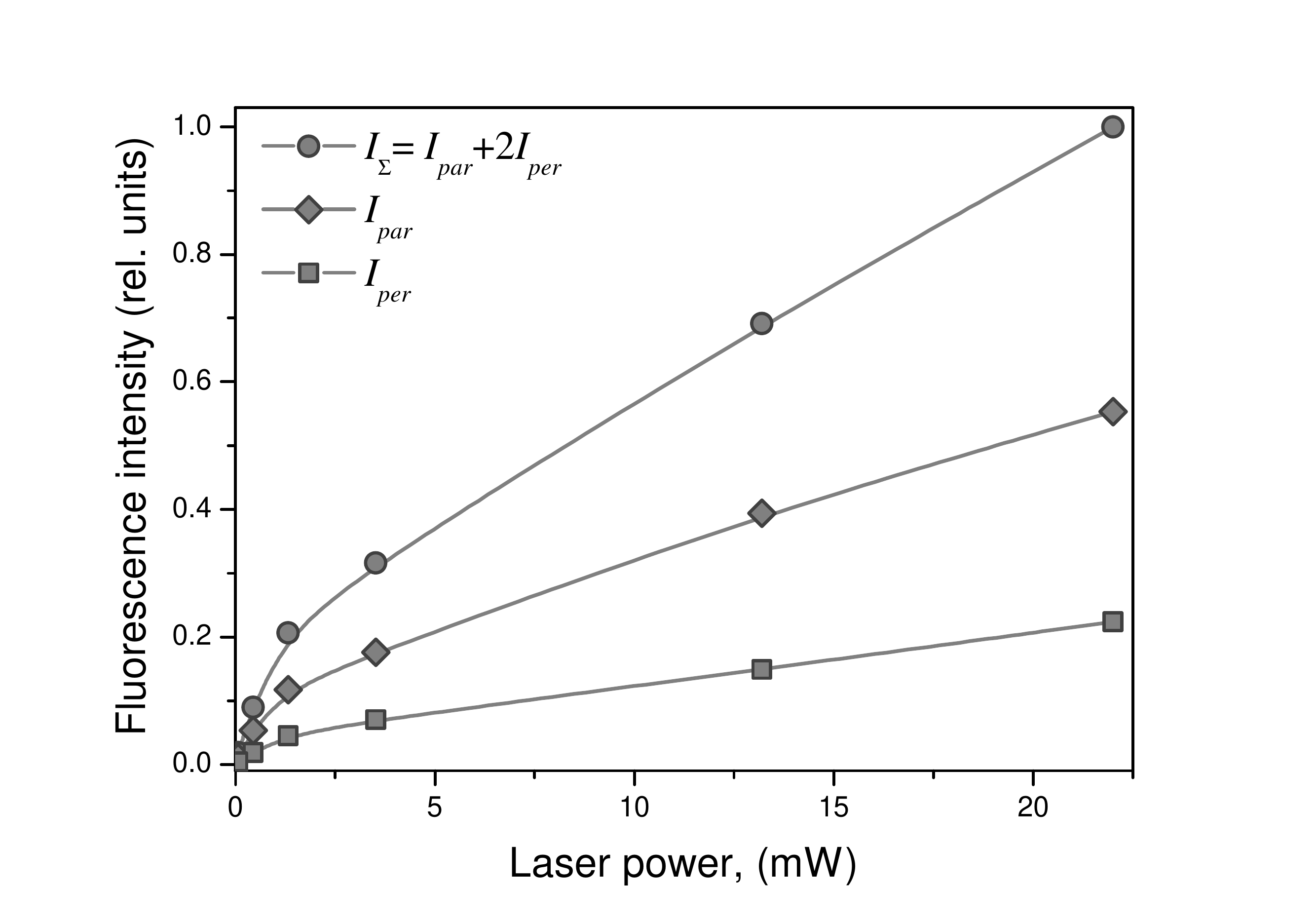}}
       \caption{\label{fig:Zeeman} Measured relative intensities versus laser power at the entrance of the cell. Here, $B=0$; the other experimental parameters are the same as in Fig.~\ref{fig:Q3}. The solid lines are shown to guide the eye.}
\end{figure}

The results presented above show that our theoretical model describes reasonably well a wide range of experimental data using only two fitting parameters: $\Omega_{R}$ and  $\gamma$. It could therefore be expected that the best fit by our model can provide estimates of relaxation and molecule-light interaction parameters that correspond to the real experimental conditions. As the shapes of magneto-optical resonances are very sensitive to these parameters, the model can be used as a tool for estimating numerically the relaxation constant $\gamma$ and the Rabi frequency $\Omega_{R}$. However, when we increased the laser power density by a factor of 360, we expected to obtain a Rabi frequency $\Omega_{R}$ that was larger by a factor of 19, see ~\eqref{Rabi}. Instead, the best fit corresponded to an increase in the Rabi frequency of only a factor of 3. The reason for this discrepancy could be that for very high laser powers, the effective power density is smaller 
than expected because of the influence of the wings of the laser beam profile, 
which has a roughly Gaussian distribution. The saturation intensity~\cite{Alnis:2003, auzinsh_book} calculated for these transitions was approximately 0.003 W/cm$^2$. The actual laser power density used in the experiment was between 0.003 and 1.1 W/cm$^2$. At high laser power densities the transition is already saturated in the most of the area that is defined as the laser beam cross section. However, the wings of the laser beam profile are not saturated and their contribution becomes larger relative to the saturated center. Therefore, the signal is being generated by a large region whose actual laser power density is much lower than the defined total laser power density of the beam. This interpretation is supported by Fig.~\ref{fig:Zeeman}, which shows that the LIF signal continues to increase 
almost linearly even though at such laser power density at the central part of the beam the transition is already fully saturated. A theoretical model accounting for a realistic Gaussian-shaped laser beam profile is computationally too intense to be applied to the molecular system discussed in this paper. Therefore, we have chosen a simpler approach that employs an effective Rabi frequency.

It is worth discussing in more detail the unexpected narrowing of the nonlinear ground state magnetic resonances in the \textit{B}-dependences of the orthogonally polarized LIF intensity \textit{I}$_{per}$ when compared to the corresponding dependences of the parallel polarized LIF component \textit{I}$_{par}$.  Such an effect is theoretically predicted, under certain conditions, by simulations using the present model and was experimentally confirmed (see Figs.~\ref{fig:Q5} and~\ref{fig:Q3}). Indeed, the half width $B_{1/2}$ of the growing part (dark resonance) of the \textit{I}$_{per}(B)$ was about $B_{1/2}\sim\ 0.05$ T in Fig.~\ref{fig:Q5}, which was much smaller than the respective half width in the  \textit{I}$_{par}(B)$ of about 0.2 T. Interestingly enough, if the growing part of the narrow dark resonances in \textit{I}$_{per}(B)$ in Fig.~\ref{fig:Q5} and Fig.~\ref{fig:Q3} is approximated roughly by a Lorentz-type dependence, when the respective Larmor frequency $\omega_{g_{i}g_{j}}(B_{1/2})=\omega_{L}$ of the precession of the ground-state magnetic moments is connected with the ground-state relaxation rate as $\gamma=|g_{i}-g_{j}|\omega_{L}$, then the respective coherence between Zeeman sublevels of $J''$ is close to $|g_{i}-g_{j}|= 4$. 
This corresponds to magnetic-field-induced relaxation of the polarization moments $\rho_{q}^{\kappa}$, which are the coefficients of expansion of the density matrix $\rho_{g_{i}g_{j}}$~\eqref{rate1} over the irreducible tensor operators (called polarization operators) with rank $\kappa=4$ (see, for instance,~\cite{auzinsh_book, Molecules} and references therein). Such a polarization moment is called a hexadecapole moment, and its manifestation in LIF is, as a rule, masked by a second-rank $\kappa=2$ quadrupole polarization moment, or alignment. The latter should produce roughly a two times broader \textit{B}-dependence, with $B_{1/2}$ corresponding to $\gamma=2\omega_{L}$. For essentially non-linear conditions, as in Fig.~\ref{fig:Q5} and Fig.~\ref{fig:Q3}, the \textit{B}-dependence of the LIF is expected to be power broadened, and such broadening should be much more pronounced in the \textit{I}$_{par}(B)$ determined by the second-rank polarization moment (alignment) than for the \textit{I}$_{per}(B)$ predominantly determined by the fourth rank $\kappa=4$ (or higher ranks $\kappa>4$, see~\cite{Molecules}). This suggests an explanation for the differences between the half widths of \textit{I}$_{par}(B)$ and  \textit{I}$_{per}(B)$ presented in Fig.~\ref{fig:Q5} and Fig.~\ref{fig:Q3}. 

\section{\label{Conclusion:level1}Conclusion}
Nonlinear magneto-optical resonances (or nonlinear Hanle signals) have been studied for systems with large angular momentum ($J\sim$100) in K$_{2}$ molecules. The dependence on temperature and laser power density was observed and described by a theoretical model that is based on the optical Bloch equations. The model included averaging over the Doppler profile, the splitting of the magnetic sublevels in an external magnetic field, and the coherence properties of the radiation field. Dark  resonances have been theoretically predicted and experimentally detected in the LIF 
intensities $I(B)$ under linearly polarized excitation. 

The theoretical simulation predicts that for the particular molecular \textit{R}-absorption transition $J''=105 \rightarrow J'=106$ the fraction of molecules that returns to the initial $J''=105$ after spontaneous emission is too small to form a bright resonance. This prediction was confirmed experimentally in this study.

The observed and calculated Hanle resonance in the component of the LIF intensity \textit{I}$_{per}(B)$ that was polarized orthogonally to $\textbf{E}$ was markedly narrower than would have been expected from the relaxation rate $\gamma$. This allowed us to explain the narrow maximum previously observed in the degree of LIF polarization in earlier works~\cite{Ferber:1979, Molecules}.

It was demonstrated that it is possible to describe nonlinear magneto-optical resonances in systems with large angular momenta over a wide range of experimental conditions with the theoretical model developed in~\cite{Blushs:2004}. As a result, theoretical modeling can serve as a tool for future investigations of systems with large angular momenta and for the development of practical applications based on them. 

\begin{acknowledgments}
We would like to thank Dr. Andrey Jarmola and Dr. Florian Gahbauer for useful advices and discussions. This work has been supported by the European Social Fund under the auspices of the project ``Support for Doctoral Studies at the University of Latvia" and by project Nr. 2009/0223/1DP/1.1.1.2.0/
09/APIA/VIAA/008. 

The support from ERAF project Nr. 2DP/2.1.1.1.0/10/APIA/VIAA/036, State Research Program Grant No. 2010/10-4/VPP-2/1 ``Development of innovative multifunctional materials, signal processing and information technology for competitive, science intensive products" and from the Latvian Science Council Grant No. 09.1036 is gratefully acknowledged.

\end{acknowledgments}

\bibliography{potassium}

\begin{thebibliography}{37}%
\makeatletter
\providecommand \@ifxundefined [1]{%
 \@ifx{#1\undefined}
}%
\providecommand \@ifnum [1]{%
 \ifnum #1\expandafter \@firstoftwo
 \else \expandafter \@secondoftwo
 \fi
}%
\providecommand \@ifx [1]{%
 \ifx #1\expandafter \@firstoftwo
 \else \expandafter \@secondoftwo
 \fi
}%
\providecommand \natexlab [1]{#1}%
\providecommand \enquote  [1]{``#1''}%
\providecommand \bibnamefont  [1]{#1}%
\providecommand \bibfnamefont [1]{#1}%
\providecommand \citenamefont [1]{#1}%
\providecommand \href@noop [0]{\@secondoftwo}%
\providecommand \href [0]{\begingroup \@sanitize@url \@href}%
\providecommand \@href[1]{\@@startlink{#1}\@@href}%
\providecommand \@@href[1]{\endgroup#1\@@endlink}%
\providecommand \@sanitize@url [0]{\catcode `\\12\catcode `\$12\catcode
  `\&12\catcode `\#12\catcode `\^12\catcode `\_12\catcode `\%12\relax}%
\providecommand \@@startlink[1]{}%
\providecommand \@@endlink[0]{}%
\providecommand \url  [0]{\begingroup\@sanitize@url \@url }%
\providecommand \@url [1]{\endgroup\@href {#1}{\urlprefix }}%
\providecommand \urlprefix  [0]{URL }%
\providecommand \Eprint [0]{\href }%
\providecommand \doibase [0]{http://dx.doi.org/}%
\providecommand \selectlanguage [0]{\@gobble}%
\providecommand \bibinfo  [0]{\@secondoftwo}%
\providecommand \bibfield  [0]{\@secondoftwo}%
\providecommand \translation [1]{[#1]}%
\providecommand \BibitemOpen [0]{}%
\providecommand \bibitemStop [0]{}%
\providecommand \bibitemNoStop [0]{.\EOS\space}%
\providecommand \EOS [0]{\spacefactor3000\relax}%
\providecommand \BibitemShut  [1]{\csname bibitem#1\endcsname}%
\let\auto@bib@innerbib\@empty
\bibitem [{\citenamefont {Hanle}(1924)}]{Hanle}%
  \BibitemOpen
  \bibfield  {author} {\bibinfo {author} {\bibfnamefont {W.}~\bibnamefont
  {Hanle}},\ }\href@noop {} {\bibfield  {journal} {\bibinfo  {journal}
  {Zeitschrift f{\"u}r Physik}\ }\textbf {\bibinfo {volume} {30}},\ \bibinfo
  {pages} {93} (\bibinfo {year} {1924})}\BibitemShut {NoStop}%
\bibitem [{\citenamefont {Kastler}(1967)}]{Kastler}%
  \BibitemOpen
  \bibfield  {author} {\bibinfo {author} {\bibfnamefont {A.}~\bibnamefont
  {Kastler}},\ }\href@noop {} {\bibfield  {journal} {\bibinfo  {journal}
  {Science}\ }\textbf {\bibinfo {volume} {158}},\ \bibinfo {pages} {214}
  (\bibinfo {year} {1967})}\BibitemShut {NoStop}%
\bibitem [{\citenamefont {Lehmann}\ and\ \citenamefont
  {Cohen-Tannoudji}(1964)}]{Lehmann:1964}%
  \BibitemOpen
  \bibfield  {author} {\bibinfo {author} {\bibfnamefont {J.~C.}\ \bibnamefont
  {Lehmann}}\ and\ \bibinfo {author} {\bibfnamefont {C.}~\bibnamefont
  {Cohen-Tannoudji}},\ }\href@noop {} {\bibfield  {journal} {\bibinfo
  {journal} {C.R. Acad. Sci. (Paris)}\ }\textbf {\bibinfo {volume} {258}},\
  \bibinfo {pages} {4463} (\bibinfo {year} {1964})}\BibitemShut {NoStop}%
\bibitem [{\citenamefont {Moruzzi}\ and\ \citenamefont
  {Strumia}(1991)}]{Strumia}%
  \BibitemOpen
  \bibfield  {author} {\bibinfo {author} {\bibfnamefont {G.}~\bibnamefont
  {Moruzzi}}\ and\ \bibinfo {author} {\bibfnamefont {F.}~\bibnamefont
  {Strumia}},\ }\href@noop {} {\emph {\bibinfo {title} {The Hanle effect and
  level-crossing spectroscopy}}},\ Physics of atoms and molecules\ (\bibinfo
  {publisher} {Plenum Press},\ \bibinfo {address} {New York},\ \bibinfo {year}
  {1991})\BibitemShut {NoStop}%
\bibitem [{\citenamefont {Budker}\ \emph {et~al.}(2002)\citenamefont {Budker},
  \citenamefont {Gawlik}, \citenamefont {Kimball}, \citenamefont {Rochester},
  \citenamefont {Yashchuk},\ and\ \citenamefont {Weis}}]{Budker:2002}%
  \BibitemOpen
  \bibfield  {author} {\bibinfo {author} {\bibfnamefont {D.}~\bibnamefont
  {Budker}}, \bibinfo {author} {\bibfnamefont {W.}~\bibnamefont {Gawlik}},
  \bibinfo {author} {\bibfnamefont {D.~F.}\ \bibnamefont {Kimball}}, \bibinfo
  {author} {\bibfnamefont {S.~M.}\ \bibnamefont {Rochester}}, \bibinfo {author}
  {\bibfnamefont {V.~V.}\ \bibnamefont {Yashchuk}}, \ and\ \bibinfo {author}
  {\bibfnamefont {A.}~\bibnamefont {Weis}},\ }\href {\doibase
  10.1103/RevModPhys.74.1153} {\bibfield  {journal} {\bibinfo  {journal}
  {Reviews of Modern Physics}\ }\textbf {\bibinfo {volume} {74}},\ \bibinfo
  {pages} {1153} (\bibinfo {year} {2002})}\BibitemShut {NoStop}%
\bibitem [{\citenamefont {Alexandrov}\ \emph {et~al.}(2005)\citenamefont
  {Alexandrov}, \citenamefont {Auzinsh}, \citenamefont {Budker}, \citenamefont
  {Kimball}, \citenamefont {Rochester},\ and\ \citenamefont
  {Yashchuk}}]{Alexandrov:2005}%
  \BibitemOpen
  \bibfield  {author} {\bibinfo {author} {\bibfnamefont {E.~B.}\ \bibnamefont
  {Alexandrov}}, \bibinfo {author} {\bibfnamefont {M.}~\bibnamefont {Auzinsh}},
  \bibinfo {author} {\bibfnamefont {D.}~\bibnamefont {Budker}}, \bibinfo
  {author} {\bibfnamefont {D.~F.}\ \bibnamefont {Kimball}}, \bibinfo {author}
  {\bibfnamefont {S.}~\bibnamefont {Rochester}}, \ and\ \bibinfo {author}
  {\bibfnamefont {V.~V.}\ \bibnamefont {Yashchuk}},\ }\href@noop {} {\bibfield
  {journal} {\bibinfo  {journal} {J. Opt. Soc. Am. B}\ }\textbf {\bibinfo
  {volume} {22}},\ \bibinfo {pages} {7} (\bibinfo {year} {2005})}\BibitemShut
  {NoStop}%
\bibitem [{\citenamefont {Dawes}\ \emph {et~al.}(2005)\citenamefont {Dawes},
  \citenamefont {Illing}, \citenamefont {Clark},\ and\ \citenamefont
  {Gauthier}}]{dawes:2005}%
  \BibitemOpen
  \bibfield  {author} {\bibinfo {author} {\bibfnamefont {A.}~\bibnamefont
  {Dawes}}, \bibinfo {author} {\bibfnamefont {L.}~\bibnamefont {Illing}},
  \bibinfo {author} {\bibfnamefont {S.}~\bibnamefont {Clark}}, \ and\ \bibinfo
  {author} {\bibfnamefont {D.}~\bibnamefont {Gauthier}},\ }\href@noop {}
  {\bibfield  {journal} {\bibinfo  {journal} {Science}\ }\textbf {\bibinfo
  {volume} {308}},\ \bibinfo {pages} {672} (\bibinfo {year}
  {2005})}\BibitemShut {NoStop}%
\bibitem [{\citenamefont {Groeger}\ \emph {et~al.}(2006)\citenamefont
  {Groeger}, \citenamefont {Bison}, \citenamefont {Knowles}, \citenamefont
  {Wynands},\ and\ \citenamefont {Weis}}]{Groeger:2006}%
  \BibitemOpen
  \bibfield  {author} {\bibinfo {author} {\bibfnamefont {S.}~\bibnamefont
  {Groeger}}, \bibinfo {author} {\bibfnamefont {G.}~\bibnamefont {Bison}},
  \bibinfo {author} {\bibfnamefont {P.}~\bibnamefont {Knowles}}, \bibinfo
  {author} {\bibfnamefont {R.}~\bibnamefont {Wynands}}, \ and\ \bibinfo
  {author} {\bibfnamefont {A.}~\bibnamefont {Weis}},\ }\href {\doibase DOI:
  10.1016/j.sna.2005.09.036} {\bibfield  {journal} {\bibinfo  {journal}
  {Sensors and Actuators A: Physical}\ }\textbf {\bibinfo {volume} {129}},\
  \bibinfo {pages} {1} (\bibinfo {year} {2006})},\ \bibinfo {note} {eMSA 2004 -
  Selected Papers from the 5th European Magnetic Sensors \& Actuators
  Conference - EMSA 2004, Cardiff, UK, 4-6 July 2004.}\BibitemShut {Stop}%
\bibitem [{\citenamefont {Budker}\ and\ \citenamefont
  {Romalis}(2007)}]{Budker:2007}%
  \BibitemOpen
  \bibfield  {author} {\bibinfo {author} {\bibfnamefont {D.}~\bibnamefont
  {Budker}}\ and\ \bibinfo {author} {\bibfnamefont {M.~V.}\ \bibnamefont
  {Romalis}},\ }\href@noop {} {\bibfield  {journal} {\bibinfo  {journal}
  {Nature Physics}\ }\textbf {\bibinfo {volume} {3}},\ \bibinfo {pages} {227}
  (\bibinfo {year} {2007})}\BibitemShut {NoStop}%
\bibitem [{\citenamefont {Lukin}\ \emph {et~al.}(1999)\citenamefont {Lukin},
  \citenamefont {Yelin}, \citenamefont {Fleischhauer},\ and\ \citenamefont
  {Scully}}]{Lukin:1999}%
  \BibitemOpen
  \bibfield  {author} {\bibinfo {author} {\bibfnamefont {M.~D.}\ \bibnamefont
  {Lukin}}, \bibinfo {author} {\bibfnamefont {S.~F.}\ \bibnamefont {Yelin}},
  \bibinfo {author} {\bibfnamefont {M.}~\bibnamefont {Fleischhauer}}, \ and\
  \bibinfo {author} {\bibfnamefont {M.~O.}\ \bibnamefont {Scully}},\ }\href
  {\doibase 10.1103/PhysRevA.60.3225} {\bibfield  {journal} {\bibinfo
  {journal} {Phys. Rev. A}\ }\textbf {\bibinfo {volume} {60}},\ \bibinfo
  {pages} {3225} (\bibinfo {year} {1999})}\BibitemShut {NoStop}%
\bibitem [{\citenamefont {Alnis}\ and\ \citenamefont
  {Auzinsh}(2001{\natexlab{a}})}]{Alnis:2001}%
  \BibitemOpen
  \bibfield  {author} {\bibinfo {author} {\bibfnamefont {J.}~\bibnamefont
  {Alnis}}\ and\ \bibinfo {author} {\bibfnamefont {M.}~\bibnamefont
  {Auzinsh}},\ }\href {\doibase 10.1103/PhysRevA.63.023407} {\bibfield
  {journal} {\bibinfo  {journal} {Physical Review A}\ }\textbf {\bibinfo
  {volume} {63}},\ \bibinfo {pages} {023407} (\bibinfo {year}
  {2001}{\natexlab{a}})},\ \Eprint {http://arxiv.org/abs/arXiv:physics/0011050}
  {arXiv:physics/0011050} \BibitemShut {NoStop}%
\bibitem [{\citenamefont {Dancheva}\ \emph {et~al.}(2000)\citenamefont
  {Dancheva}, \citenamefont {Alzetta}, \citenamefont {Cartalava}, \citenamefont
  {Taslakov},\ and\ \citenamefont {Andreeva}}]{Dancheva:2000}%
  \BibitemOpen
  \bibfield  {author} {\bibinfo {author} {\bibfnamefont {Y.}~\bibnamefont
  {Dancheva}}, \bibinfo {author} {\bibfnamefont {G.}~\bibnamefont {Alzetta}},
  \bibinfo {author} {\bibfnamefont {S.}~\bibnamefont {Cartalava}}, \bibinfo
  {author} {\bibfnamefont {M.}~\bibnamefont {Taslakov}}, \ and\ \bibinfo
  {author} {\bibfnamefont {C.}~\bibnamefont {Andreeva}},\ }\href@noop {}
  {\bibfield  {journal} {\bibinfo  {journal} {Optics Communications}\ }\textbf
  {\bibinfo {volume} {178}},\ \bibinfo {pages} {103} (\bibinfo {year}
  {2000})}\BibitemShut {NoStop}%
\bibitem [{\citenamefont {Alzetta}\ \emph {et~al.}(1976)\citenamefont
  {Alzetta}, \citenamefont {Gozzini}, \citenamefont {Moi},\ and\ \citenamefont
  {Orriols}}]{Alzetta:1976}%
  \BibitemOpen
  \bibfield  {author} {\bibinfo {author} {\bibfnamefont {G.}~\bibnamefont
  {Alzetta}}, \bibinfo {author} {\bibfnamefont {A.}~\bibnamefont {Gozzini}},
  \bibinfo {author} {\bibfnamefont {L.}~\bibnamefont {Moi}}, \ and\ \bibinfo
  {author} {\bibfnamefont {G.}~\bibnamefont {Orriols}},\ }\href@noop {}
  {\bibfield  {journal} {\bibinfo  {journal} {Il Nuovo Cimento B}\ }\textbf
  {\bibinfo {volume} {36}},\ \bibinfo {pages} {5} (\bibinfo {year}
  {1976})}\BibitemShut {NoStop}%
\bibitem [{\citenamefont {Arimondo}(1996)}]{Arimondo:1996}%
  \BibitemOpen
  \bibfield  {author} {\bibinfo {author} {\bibfnamefont {E.}~\bibnamefont
  {Arimondo}},\ }\href {http://www.worldcat.org/isbn/0444823093} {\emph
  {\bibinfo {title} {Progress in Optics, Vol. 35}}},\ edited by\ \bibinfo
  {editor} {\bibfnamefont {D.}~\bibnamefont {Alsina}},\ Vol.~\bibinfo {volume}
  {35}\ (\bibinfo  {publisher} {Elsevier},\ \bibinfo {year} {1996})\ pp.\
  \bibinfo {pages} {257--354}\BibitemShut {NoStop}%
\bibitem [{\citenamefont {Fleischhauer}\ \emph {et~al.}(2005)\citenamefont
  {Fleischhauer}, \citenamefont {Imamoglu},\ and\ \citenamefont
  {Marangos}}]{Fleisch:2005}%
  \BibitemOpen
  \bibfield  {author} {\bibinfo {author} {\bibfnamefont {M.}~\bibnamefont
  {Fleischhauer}}, \bibinfo {author} {\bibfnamefont {A.}~\bibnamefont
  {Imamoglu}}, \ and\ \bibinfo {author} {\bibfnamefont {J.~P.}\ \bibnamefont
  {Marangos}},\ }\href {\doibase 10.1103/RevModPhys.77.633} {\bibfield
  {journal} {\bibinfo  {journal} {Reviews of Modern Physics}\ }\textbf
  {\bibinfo {volume} {77}},\ \bibinfo {pages} {633} (\bibinfo {year}
  {2005})}\BibitemShut {NoStop}%
\bibitem [{\citenamefont {Budker}\ and\ \citenamefont
  {Rochester}(2004)}]{Budker:2004}%
  \BibitemOpen
  \bibfield  {author} {\bibinfo {author} {\bibfnamefont {D.}~\bibnamefont
  {Budker}}\ and\ \bibinfo {author} {\bibfnamefont {S.~M.}\ \bibnamefont
  {Rochester}},\ }\href@noop {} {\bibfield  {journal} {\bibinfo  {journal}
  {Physical Review A}\ }\textbf {\bibinfo {volume} {70}},\ \bibinfo {pages}
  {025804} (\bibinfo {year} {2004})}\BibitemShut {NoStop}%
\bibitem [{\citenamefont {Hau}\ \emph {et~al.}(1999)\citenamefont {Hau},
  \citenamefont {Harris}, \citenamefont {Dutton},\ and\ \citenamefont
  {Behroozi}}]{Hau:1999}%
  \BibitemOpen
  \bibfield  {author} {\bibinfo {author} {\bibfnamefont {L.~V.}\ \bibnamefont
  {Hau}}, \bibinfo {author} {\bibfnamefont {S.~E.}\ \bibnamefont {Harris}},
  \bibinfo {author} {\bibfnamefont {Z.}~\bibnamefont {Dutton}}, \ and\ \bibinfo
  {author} {\bibfnamefont {C.~H.}\ \bibnamefont {Behroozi}},\ }\href {\doibase
  10.1038/17561} {\bibfield  {journal} {\bibinfo  {journal} {Nature}\ }\textbf
  {\bibinfo {volume} {397}},\ \bibinfo {pages} {594} (\bibinfo {year}
  {1999})}\BibitemShut {NoStop}%
\bibitem [{\citenamefont {Auzinsh}\ and\ \citenamefont
  {Ferber}(1995)}]{Molecules}%
  \BibitemOpen
  \bibfield  {author} {\bibinfo {author} {\bibfnamefont {M.}~\bibnamefont
  {Auzinsh}}\ and\ \bibinfo {author} {\bibfnamefont {R.}~\bibnamefont
  {Ferber}},\ }\href@noop {} {\emph {\bibinfo {title} {Optical Polarization of
  Molecules}}}\ (\bibinfo  {publisher} {Cambridge University Press},\ \bibinfo
  {year} {1995})\BibitemShut {NoStop}%
\bibitem [{\citenamefont {Papoyan}\ \emph {et~al.}(2002)\citenamefont
  {Papoyan}, \citenamefont {Auzinsh},\ and\ \citenamefont
  {Bergmann.}}]{Papoyan:2002}%
  \BibitemOpen
  \bibfield  {author} {\bibinfo {author} {\bibfnamefont {A.~V.}\ \bibnamefont
  {Papoyan}}, \bibinfo {author} {\bibfnamefont {M.}~\bibnamefont {Auzinsh}}, \
  and\ \bibinfo {author} {\bibfnamefont {K.}~\bibnamefont {Bergmann.}},\
  }\href@noop {} {\bibfield  {journal} {\bibinfo  {journal} {European Physical
  Journal D}\ }\textbf {\bibinfo {volume} {21}},\ \bibinfo {pages} {63}
  (\bibinfo {year} {2002})}\BibitemShut {NoStop}%
\bibitem [{\citenamefont {Auzinsh}\ \emph {et~al.}(2010)\citenamefont
  {Auzinsh}, \citenamefont {Budker},\ and\ \citenamefont
  {Rochester}}]{auzinsh_book}%
  \BibitemOpen
  \bibfield  {author} {\bibinfo {author} {\bibfnamefont {M.}~\bibnamefont
  {Auzinsh}}, \bibinfo {author} {\bibfnamefont {D.}~\bibnamefont {Budker}}, \
  and\ \bibinfo {author} {\bibfnamefont {S.~M.}\ \bibnamefont {Rochester}},\
  }\href@noop {} {\emph {\bibinfo {title} {Optically polarized atoms:
  understanding light-atom interactions}}}\ (\bibinfo  {publisher} {Oxford
  University Press, USA},\ \bibinfo {year} {2010})\BibitemShut {NoStop}%
\bibitem [{\citenamefont {Broyer}\ \emph {et~al.}(1973)\citenamefont {Broyer},
  \citenamefont {Dalby}, \citenamefont {Vigu{\'e}},\ and\ \citenamefont
  {Lehmann}}]{Broyer:1973}%
  \BibitemOpen
  \bibfield  {author} {\bibinfo {author} {\bibfnamefont {M.}~\bibnamefont
  {Broyer}}, \bibinfo {author} {\bibfnamefont {F.~W.}\ \bibnamefont {Dalby}},
  \bibinfo {author} {\bibfnamefont {J.}~\bibnamefont {Vigu{\'e}}}, \ and\
  \bibinfo {author} {\bibfnamefont {J.~C.}\ \bibnamefont {Lehmann}},\
  }\href@noop {} {\bibfield  {journal} {\bibinfo  {journal} {Can. J. Phys.}\ }
  (\bibinfo {year} {1973})}\BibitemShut {NoStop}%
\bibitem [{\citenamefont {Solarz}\ and\ \citenamefont
  {Levy}(1972)}]{Solarz:1972}%
  \BibitemOpen
  \bibfield  {author} {\bibinfo {author} {\bibfnamefont {R.}~\bibnamefont
  {Solarz}}\ and\ \bibinfo {author} {\bibfnamefont {D.~H.}\ \bibnamefont
  {Levy}},\ }\href {\doibase DOI: 10.1016/0009-2614(72)80317-8} {\bibfield
  {journal} {\bibinfo  {journal} {Chemical Physics Letters}\ }\textbf {\bibinfo
  {volume} {17}},\ \bibinfo {pages} {35} (\bibinfo {year} {1972})}\BibitemShut
  {NoStop}%
\bibitem [{\citenamefont {Ferber}\ \emph {et~al.}(1979)\citenamefont {Ferber},
  \citenamefont {Shmit},\ and\ \citenamefont {Tamanis}}]{Ferber:1979}%
  \BibitemOpen
  \bibfield  {author} {\bibinfo {author} {\bibfnamefont {R.~S.}\ \bibnamefont
  {Ferber}}, \bibinfo {author} {\bibfnamefont {O.~A.}\ \bibnamefont {Shmit}}, \
  and\ \bibinfo {author} {\bibfnamefont {M.~Y.}\ \bibnamefont {Tamanis}},\
  }\href {\doibase DOI: 10.1016/0009-2614(79)87145-6} {\bibfield  {journal}
  {\bibinfo  {journal} {Chemical Physics Letters}\ }\textbf {\bibinfo {volume}
  {61}},\ \bibinfo {pages} {441} (\bibinfo {year} {1979})}\BibitemShut
  {NoStop}%
\bibitem [{\citenamefont {Drullinger}\ and\ \citenamefont
  {Zare}(1969)}]{Drullinger:1969}%
  \BibitemOpen
  \bibfield  {author} {\bibinfo {author} {\bibfnamefont {R.~E.}\ \bibnamefont
  {Drullinger}}\ and\ \bibinfo {author} {\bibfnamefont {R.~N.}\ \bibnamefont
  {Zare}},\ }\href@noop {} {\bibfield  {journal} {\bibinfo  {journal} {J. Chem.
  Phys.}\ }\textbf {\bibinfo {volume} {51}},\ \bibinfo {pages} {5532} (\bibinfo
  {year} {1969})}\BibitemShut {NoStop}%
\bibitem [{\citenamefont {Ducloy}(1976)}]{Ducloy:1976}%
  \BibitemOpen
  \bibfield  {author} {\bibinfo {author} {\bibfnamefont {M.}~\bibnamefont
  {Ducloy}},\ }\href {http://stacks.iop.org/0022-3700/9/i=3/a=005} {\bibfield
  {journal} {\bibinfo  {journal} {Journal of Physics B: Atomic and Molecular
  Physics}\ }\textbf {\bibinfo {volume} {9}},\ \bibinfo {pages} {357} (\bibinfo
  {year} {1976})}\BibitemShut {NoStop}%
\bibitem [{\citenamefont {Auzinsh}\ and\ \citenamefont
  {Ferber}(1991)}]{Auzinsh:1991}%
  \BibitemOpen
  \bibfield  {author} {\bibinfo {author} {\bibfnamefont {M.~P.}\ \bibnamefont
  {Auzinsh}}\ and\ \bibinfo {author} {\bibfnamefont {R.~S.}\ \bibnamefont
  {Ferber}},\ }\href {\doibase 10.1103/PhysRevA.43.2374} {\bibfield  {journal}
  {\bibinfo  {journal} {Phys. Rev. A}\ }\textbf {\bibinfo {volume} {43}},\
  \bibinfo {pages} {2374} (\bibinfo {year} {1991})}\BibitemShut {NoStop}%
\bibitem [{\citenamefont {Renzoni}\ \emph {et~al.}(2001)\citenamefont
  {Renzoni}, \citenamefont {Cartaleva}, \citenamefont {Alzetta},\ and\
  \citenamefont {Arimondo}}]{Renzoni:2001a}%
  \BibitemOpen
  \bibfield  {author} {\bibinfo {author} {\bibfnamefont {F.}~\bibnamefont
  {Renzoni}}, \bibinfo {author} {\bibfnamefont {S.}~\bibnamefont {Cartaleva}},
  \bibinfo {author} {\bibfnamefont {G.}~\bibnamefont {Alzetta}}, \ and\
  \bibinfo {author} {\bibfnamefont {E.}~\bibnamefont {Arimondo}},\ }\href@noop
  {} {\bibfield  {journal} {\bibinfo  {journal} {Physical Review A}\ }\textbf
  {\bibinfo {volume} {63}},\ \bibinfo {pages} {065401} (\bibinfo {year}
  {2001})}\BibitemShut {NoStop}%
\bibitem [{\citenamefont {Alnis}\ and\ \citenamefont
  {Auzinsh}(2001{\natexlab{b}})}]{AlnisJPB:2001}%
  \BibitemOpen
  \bibfield  {author} {\bibinfo {author} {\bibfnamefont {J.}~\bibnamefont
  {Alnis}}\ and\ \bibinfo {author} {\bibfnamefont {M.}~\bibnamefont
  {Auzinsh}},\ }\href@noop {} {\bibfield  {journal} {\bibinfo  {journal}
  {Journal of Physics B Atomic Molecular Physics}\ }\textbf {\bibinfo {volume}
  {34}},\ \bibinfo {pages} {3889} (\bibinfo {year} {2001}{\natexlab{b}})},\
  \Eprint {http://arxiv.org/abs/arXiv:physics/0011051} {arXiv:physics/0011051}
  \BibitemShut {NoStop}%
\bibitem [{\citenamefont {Auzinsh}\ \emph {et~al.}(2009)\citenamefont
  {Auzinsh}, \citenamefont {Ferber}, \citenamefont {Gahbauer}, \citenamefont
  {Jarmola},\ and\ \citenamefont {Kalvans}}]{Auzinsh:2009}%
  \BibitemOpen
  \bibfield  {author} {\bibinfo {author} {\bibfnamefont {M.}~\bibnamefont
  {Auzinsh}}, \bibinfo {author} {\bibfnamefont {R.}~\bibnamefont {Ferber}},
  \bibinfo {author} {\bibfnamefont {F.}~\bibnamefont {Gahbauer}}, \bibinfo
  {author} {\bibfnamefont {A.}~\bibnamefont {Jarmola}}, \ and\ \bibinfo
  {author} {\bibfnamefont {L.}~\bibnamefont {Kalvans}},\ }\href {\doibase
  10.1103/PhysRevA.79.053404} {\bibfield  {journal} {\bibinfo  {journal}
  {Physical Review A}\ }\textbf {\bibinfo {volume} {79}},\ \bibinfo {pages}
  {053404} (\bibinfo {year} {2009})}\BibitemShut {NoStop}%
\bibitem [{\citenamefont {Blushs}\ and\ \citenamefont
  {Auzinsh}(2004)}]{Blushs:2004}%
  \BibitemOpen
  \bibfield  {author} {\bibinfo {author} {\bibfnamefont {K.}~\bibnamefont
  {Blushs}}\ and\ \bibinfo {author} {\bibfnamefont {M.}~\bibnamefont
  {Auzinsh}},\ }\href@noop {} {\bibfield  {journal} {\bibinfo  {journal}
  {Physical Review A}\ }\textbf {\bibinfo {volume} {69}},\ \bibinfo {pages}
  {063806} (\bibinfo {year} {2004})}\BibitemShut {NoStop}%
\bibitem [{\citenamefont {Stenholm}(2005)}]{Stenholm:2005}%
  \BibitemOpen
  \bibfield  {author} {\bibinfo {author} {\bibfnamefont {S.}~\bibnamefont
  {Stenholm}},\ }\href@noop {} {\emph {\bibinfo {title} {Foundations of Laser
  Spectroscopy}}}\ (\bibinfo {address} {Dover Publications, Inc., Mineoloa, New
  York},\ \bibinfo {year} {2005})\BibitemShut {NoStop}%
\bibitem [{\citenamefont {Allen}\ and\ \citenamefont
  {Eberly}(1975)}]{Allen:1975}%
  \BibitemOpen
  \bibfield  {author} {\bibinfo {author} {\bibfnamefont {L.}~\bibnamefont
  {Allen}}\ and\ \bibinfo {author} {\bibfnamefont {J.~H.}\ \bibnamefont
  {Eberly}},\ }\href@noop {} {\emph {\bibinfo {title} {Optical Resonance and
  Two Level Atoms}}}\ (\bibinfo {address} {Wiley, New York},\ \bibinfo {year}
  {1975})\BibitemShut {NoStop}%
\bibitem [{\citenamefont {van Kampen}(1976)}]{Kampen:1976}%
  \BibitemOpen
  \bibfield  {author} {\bibinfo {author} {\bibfnamefont {N.~G.}\ \bibnamefont
  {van Kampen}},\ }\href@noop {} {\bibfield  {journal} {\bibinfo  {journal}
  {Physics Reports}\ }\textbf {\bibinfo {volume} {24}},\ \bibinfo {pages} {171}
  (\bibinfo {year} {1976})}\BibitemShut {NoStop}%
\bibitem [{\citenamefont {Davis}(2004)}]{Davis2004_1}%
  \BibitemOpen
  \bibfield  {author} {\bibinfo {author} {\bibfnamefont {T.~A.}\ \bibnamefont
  {Davis}},\ }\href {\doibase http://doi.acm.org/10.1145/992200.992205}
  {\bibfield  {journal} {\bibinfo  {journal} {ACM Trans. Math. Softw.}\
  }\textbf {\bibinfo {volume} {30}},\ \bibinfo {pages} {165} (\bibinfo {year}
  {2004})}\BibitemShut {NoStop}%
\bibitem [{\citenamefont {Herzberg}(1950)}]{Herzberg}%
  \BibitemOpen
  \bibfield  {author} {\bibinfo {author} {\bibfnamefont {G.}~\bibnamefont
  {Herzberg}},\ }\href@noop {} {\emph {\bibinfo {title} {Molecular spectra and
  molecular structure. Vol.1: Spectra of diatomic molecules}}}\ (\bibinfo
  {publisher} {New York: Van Nostrand Reinhold},\ \bibinfo {year}
  {1950})\BibitemShut {NoStop}%
\bibitem [{\citenamefont {Nesmeyanov}(1963)}]{Nesmeyanov}%
  \BibitemOpen
  \bibfield  {author} {\bibinfo {author} {\bibfnamefont {A.~N.}\ \bibnamefont
  {Nesmeyanov}},\ }\href@noop {} {\emph {\bibinfo {title} {Vapour Preasure of
  the Elements}}},\ edited by\ \bibinfo {editor} {\bibfnamefont {J.~I.}\
  \bibnamefont {Carasso}}\ (\bibinfo  {publisher} {Academic Press, New York},\
  \bibinfo {year} {1963})\BibitemShut {NoStop}%
\bibitem [{\citenamefont {Alnis}\ \emph {et~al.}(2003)\citenamefont {Alnis},
  \citenamefont {Blushs}, \citenamefont {Auzinsh}, \citenamefont {Kennedy},
  \citenamefont {Shafer-Ray},\ and\ \citenamefont {Abraham}}]{Alnis:2003}%
  \BibitemOpen
  \bibfield  {author} {\bibinfo {author} {\bibfnamefont {J.}~\bibnamefont
  {Alnis}}, \bibinfo {author} {\bibfnamefont {K.}~\bibnamefont {Blushs}},
  \bibinfo {author} {\bibfnamefont {M.}~\bibnamefont {Auzinsh}}, \bibinfo
  {author} {\bibfnamefont {S.}~\bibnamefont {Kennedy}}, \bibinfo {author}
  {\bibfnamefont {N.}~\bibnamefont {Shafer-Ray}}, \ and\ \bibinfo {author}
  {\bibfnamefont {E.}~\bibnamefont {Abraham}},\ }\href@noop {} {\bibfield
  {journal} {\bibinfo  {journal} {Journal of Physics B}\ }\textbf {\bibinfo
  {volume} {36}},\ \bibinfo {pages} {1161} (\bibinfo {year}
  {2003})}\BibitemShut {NoStop}%
\end{thebibliography}%

\end{document}